# Topological Compactons *

H. Arodź

Marian Smoluchowski Institute of Physics, Jagellonian University,
Reymonta 4, 30-059 Cracow, Poland

**Abstract**

One dimensional topological kink which has strictly finite size without any exponential or power-like tail is presented. It can be observed in a simple mechanical system akin to the one used in order to demonstrate sinus-Gordon solitons.

Preprint TPJU-13/2001
---
*Paper supported in part by ESF "COSLAB" Programme

# 1 Introduction

Extended stable or long-lived objects like kinks, vortices or domain walls are interesting for several reasons. First, they are copiously produced in condensed matter systems, especially in time dependent conditions. Second, they have non-perturbative origin, often related to existence of topologically nontrivial sectors in pertinent models. Resulting complexity of their static characteristics, of evolution and interactions poses a challenge to theorists as well as to experimenters. Third, they likely were produced in phase transitions in the early Universe and contributed to later evolution of it. Finally, physics of the extended objects inspires avant-garde ideas, for example, about the four-dimensional Universe being a domain wall. All these aspects are discussed in numerous papers. Introductory lectures and extensive lists of references can be found in, e.g., [1, 2, 3, 4, 5].

In the present paper we concentrate on topological kinks. In a typical case one describes them in terms of a field theoretical model which has a non connected vacuum manifold. Basic kink is static. The pertinent solutions of field equations smoothly interpolate between a pair of classical vacua belonging to different connected components of the vacuum manifold. Typically, the fields approach the classical vacua exponentially. From the static kink one can get a moving one by Lorentzian or Galilean boosts.

The topological kinks are particularly important because they are related to domain walls. Consider a stable planar domain wall, homogeneous in the two directions along the domain wall. The corresponding fields depend only on a coordinate $\xi$ parametrizing the direction perpendicular to the domain wall. They are precisely the same as for the kink. Thus, the kink can be identified with the transverse profile of the planar domain wall. The planar domain wall is the main prerequisite of analytical description of curved domain walls with a help of expansion in width, see, e.g., [6].

We consider a field theoretical model which involves a single, real scalar field $\phi$ with double degerate classical ground state $\pm\phi_0$. This is a typical set up for theoretical investigations of domain walls. However, in our case the field potential $V(\phi)$ is not smooth. It is merely continuous, and it steeply increases to infinity at the ground state values $\pm\phi_0$ of the field. At first glance such a model might look quite artificial. As a matter of fact, it is just the opposite — such field potential has very simple mechanical realization described in next Section. One can easily perform experiments in which the one-dimensional kinks are observed. This is first attractive feature of the presented model.

Another interesting fact has to do with the theoretical side of the model. It turns out that the kink in our model has the peculiar feature that at certain finite distance $\xi_0$ from the center of it the scalar field reaches exactly the classical ground states $\pm\phi_0$ and remains constant at larger distances. The common exponential approach to the ground state is absent here.

Solutions of non-linear evolution equations which have compact support are called compactons. They were obtained for various modified Korteweg-de Vries equations [7, 8, 9, 10, 11]. The compacton presented in our paper is distinguished by two features. First, it has a nonvanishing topological charge related to the degenerate classical ground state, and therefore its embedding in a higher dimensional space gives a stable domain wall. In the K-dV case, higher dimensional generalizations are not straightrorward. Second, our compacton can be observed in the simple mechanical system.

The plan of our paper is as follows. In Section 2 we discuss the topological compacton. The mechanical system with compactons is presented in Section 3. Section 4 contains several remarks.

## 2  The topological compacton

We consider one dimensional topological kinks. The space coordinate is denoted by $\xi$, and the time coordinate by $\tau$. Moreover, $\tau$ and $\xi$ are dimensionless. The reason for this notation is that in theoretical analysis of the mechanical model presented in next Section $\tau$ and $\xi$ appear as rescaled ordinary time $t$ and position $x$, respectively.

Lagrangian of our model has the form

$$L = \frac{1}{2}(\partial_\tau \phi)^2 - \frac{1}{2}(\partial_\xi \phi)^2 - V(\phi), \qquad (1)$$

where $V(\phi)$ is the field potential. In order to keep the following discussion simple, we assume that the ground state in the model is double degenerate, and that it is obtained for $\phi = \pm\phi_0$. The field potential is symmetric, $V(\phi) = V(-\phi)$, and it has a local maximum at $\phi = 0$. Considerations presented below can easily be generalized to fields potentials of other forms.

As always, the kinks are given by solutions of the field equation obtained from Lagrangian (1). They interpolate between the two ground state values of the field. For the static kink the field equation has the form

$$\partial_\xi^2 \phi - V'(\phi) = 0 \qquad (2)$$

where $V' = dV/d\phi$. Multiplying Eq.(2) by $\partial_\xi \phi$ and integrating over $\xi$ we obtain the equation

$$\frac{1}{2}(\partial_\xi \phi)^2 - V(\phi) = \text{constant}. \qquad (3)$$

We expect that $\partial_\xi \phi \to 0$ when $\phi \to \phi_0$. Therefore, constant $= -V(\phi_0)$, and Eq.(3) can be written in the form

$$(\partial_\xi \phi)^2 = 2\left(V(\phi) - V(\phi_0)\right). \tag{4}$$

For the kink, $\phi(\xi)$ increases with $\xi$ monotonically from $-\phi_0$ to $+\phi_0$. Hence, $\partial_\xi \phi \geq 0$, and Eq.(4) can be written as

$$\partial_\xi \phi = \sqrt{2\left(V(\phi) - V(\phi_0)\right)}. \tag{5}$$

This equation is very convenient starting point for a detailed analysis of the kink.

Let us first have a look at the behaviour of $\phi$ when it approaches $\phi_0$. Then,

$$\phi(\xi) = \phi_0 - \delta\phi(\xi),$$

where $\delta\phi \geq 0$. Expanding the l.h.s. of Eq.(5) in $\delta\phi$ and keeping only the leading term we obtain the following equation

$$\partial_\xi \delta\phi = -\sqrt{2|V'(\phi_0)|}\sqrt{\delta\phi}. \tag{6}$$

Here $V'(\phi_0)$ is of course understood as the limit from the side of $\phi < \phi_0$ (the same applies to all derivatives $V^{(k)}(\phi_0)$ below). General solution of Eq.(6) has the form

$$\delta\phi(\xi) \cong \frac{1}{2}|V'(\phi_0)|(\xi_0 - \xi)^2, \tag{7}$$

where $\xi_0$ is an arbitrary constant. Thus, we have obtained the parabolic approach to the ground state value of the field $\phi_0$. This value is reached at $\xi = \xi_0$ exactly.

The unusual polynomial approach of the field to its ground state value is of course due to the fact that $V'(\phi_0) \neq 0$. Then, in the expansion

$$V(\phi) - V(\phi_0) = V'(\phi_0)(\phi - \phi_0) \tag{8}$$
$$+ \frac{1}{2}V''(\phi_0)(\phi - \phi_0)^2 + \frac{1}{3!}V'''(\phi_0)(\phi - \phi_0)^3 + \ldots$$

the first term dominates the limit $\phi \to \phi_0-$. The well-known exponential approach is obtained when $V'(\phi_0) = 0$ and $V''(\phi_0) > 0$. In this case

$$\delta\phi(\xi) \cong \exp(-\sqrt{V''(\phi_0)}\xi). \tag{9}$$

If the first nonvanishing term is of the order $k > 2$ and $(-1)^k V^{(k)}(\phi_0) > 0$ then the approach to $\phi_0$ is power-like with a negative power of $\xi$, namely

$$\delta\phi(\xi) \cong \left[(\frac{k}{2} - 1)\sqrt{\frac{2|V^{(k)}(\phi_0)|}{k!}}\right]^{\frac{2}{2-k}} \frac{1}{\xi^{\frac{2}{k-2}}}. \tag{10}$$

The power-like asymptotics (10) is sufficient for convergence at large $\xi$ of the integral in the expression for the total energy $E_k$ of the kink,

$$E_k = \int_{-\infty}^{+\infty} d\xi \left( \frac{1}{2}(\partial_\xi \phi)^2 + V(\phi) - V(\phi_0) \right). \tag{11}$$

The term $V(\phi_0)$ is included in order to subtract the energy of the ground state.

The kink located at $\xi = 0$ is the solution $\phi(\xi)$ of Eq.(5) which has the following dependence on $\xi$: $\phi(\xi)$ is equal to $-\phi_0$ for $\xi \leq -\xi_0$, then it starts to increase in the parabolic manner,

$$\phi(\xi) \cong -\phi_0 + \frac{1}{2}|V'(\phi_0)|(\xi + \xi_0)^2,$$

it reaches 0 at $\xi = 0$, and continues to increase to $+\phi_0$ as $\xi$ approaches $\xi_0$. Close to $\xi_0$

$$\phi(\xi) \cong \phi_0 - \frac{1}{2}|V'(\phi_0)|(\xi - \xi_0)^2.$$

Finally, $\phi(\xi) = \phi_0$ for $\xi \geq \xi_0$. Anti-kink is obtained by changing the sign of $\phi(\xi)$. It is clear that such kinks and anti-kinks are compactons.

As the example, let us take the field potential $V(\phi)$ of the form

$$V(\phi) = \begin{cases} \cos\phi - 1 & \text{for} \quad |\phi| \leq \phi_0 \\ \infty & \text{for} \quad |\phi| > \phi_0, \end{cases} \tag{12}$$

where $\phi_0$ is a constant, $\pi > \phi_0 > 0$, see Fig.1. It has two degenerate minima at $\phi = \pm\phi_0$.

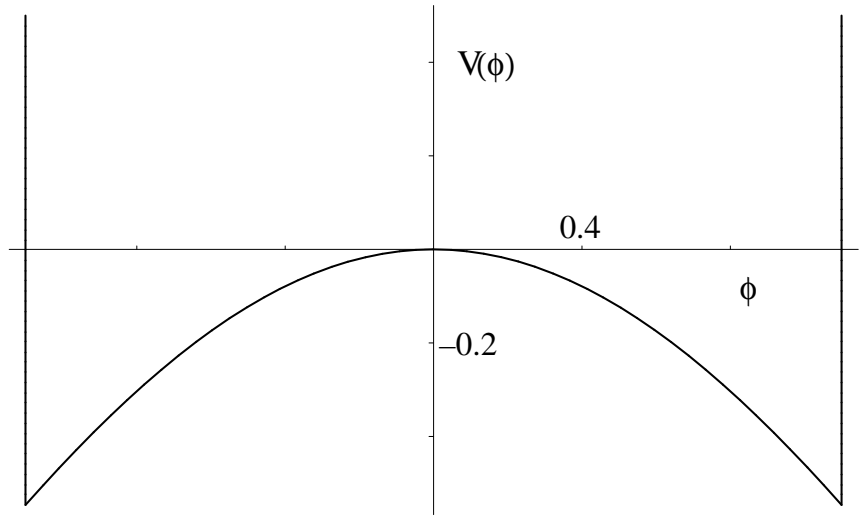

Fig.1. Plot of the potential $V(\phi)$ for $\phi_0 = 1.1$.

Because of the infinite potential barrier, modulus of the field $\phi$ cannot exceed $\phi_0$. Therefore, Eq.(5) is now supplemented by the condition

$$-\phi_0 \leq \phi \leq \phi_0. \tag{13}$$

Potential of this kind has the simple mechanical realisation described in next Section. The angle $\phi_0$ is an external parameter which we can easily regulate.

The length of the compacton is equal to $2\xi_0$. It is related to $\phi_0$ by the following formula obtained by integration of Eq.(5)

$$\int_0^{\phi_0} d\phi \frac{1}{\sqrt{\cos\phi - \cos\phi_0}} = \sqrt{2}\xi_0.$$

The change of the integration variable to $\lambda = \phi/\phi_0$ gives formula

$$\int_0^1 d\lambda \frac{1}{\sqrt{\cos(\lambda\phi_0) - \cos\phi_0}} = \frac{\sqrt{2}}{\phi_0}\xi_0, \tag{14}$$

from which we can find the dependence of $\xi_0$ on $\phi_0$. In particular, for small $\phi_0$ we may expand the cosine functions. Keeping the terms up to $\sim \phi_0^4$ we obtain the following formula

$$\xi_0 = \frac{\pi}{2}\left(1 + \frac{\phi_0^2}{16} + \ldots\right). \tag{15}$$

Thus, the length of the kink remains finite in the limit $\phi_0 \to 0$. Nevertheless, in this limit the kink of course disappears because it becomes completely flat, $\phi(\xi) \approx 0$.

Because $\phi(-\xi) = -\phi(\xi)$, it is sufficient to obtain the kink solution in the half-line $\xi \geq 0$. The initial value of $\phi$ at $\xi = 0$ is known: $\phi(0) = 0$. Then, Eq.(5) is equivalent to the following integral equation

$$\int_0^{\phi(\xi)/\phi_0} \frac{d\lambda}{\sqrt{\cos(\phi_0\lambda) - \cos\phi_0}} = \frac{\sqrt{2}}{\phi_0}\xi. \tag{16}$$

The integral on the l.h.s. can be related to an elliptic function. The integral equation (16) can be used for calculation of the approximate forms of $\phi(\xi)$. For small $\xi$ we just expand in $\lambda$ the cosine function $\cos(\phi_0\lambda)$ and the inverse of the square root. We find that for $\xi \to 0$

$$\phi(\xi) = 2\sin\frac{\phi_0}{2}\xi - \frac{1}{3}\sin\frac{\phi_0}{2}\xi^3 + \ldots. \tag{17}$$

In order to obtain $\phi(\xi)$ for $\xi \to \xi_0-$ we subtract from both sides of Eq.(16) $\sqrt{2}\xi_0/\phi_0$, and on the l.h.s. we use formula (14). We obtain the integral over $\lambda$ in the interval $[\phi(\xi)/\phi_0, 1]$. Such integral can be tackled by expanding the integrand in $\epsilon = 1 - \lambda$. Finally, we find that for $\xi \to \xi_0-$

$$\phi(\xi) = \phi_0 - \frac{\sin \phi_0}{2}(\xi - \xi_0)^2 - \frac{\sin(2\phi_0)}{48}(\xi - \xi_0)^4 + \ldots . \qquad (18)$$

Also obtaining a numerical solution is rather easy. The solution $\phi(\xi)$ starts from $\phi = 0$ at $\xi = 0$ with the slope

$$\partial_\xi \phi|_{\xi=0} = 2 \sin \frac{\phi_0}{2},$$

and it has the smooth shape presented in Fig.2. It has been obtained with the help of Maple©.

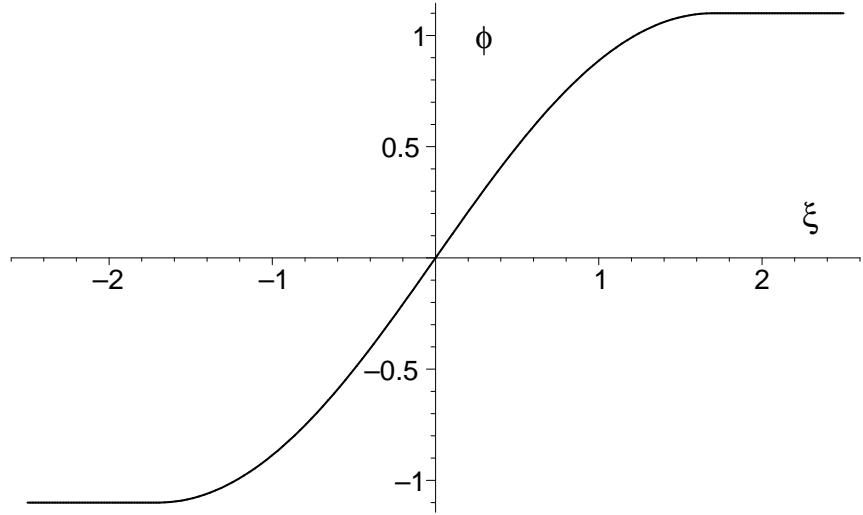

Fig.2. The numerical solution $\phi(\xi)$ of Eq.(5). $V(\phi)$ is given by formula (12) with $\phi_0 = 1.1$.

The total energy of the field is given by the formula

$$E = \int_{-\xi_0}^{\xi_0} d\xi \left[\frac{1}{2}(\partial_\xi \phi)^2 + \cos \phi - 1\right]. \qquad (19)$$

Using Eq.(4) we can write it also as

$$E = -2\xi_0(1 - \cos \phi_0) + \int_{-\xi_0}^{\xi_0} d\xi (\partial_\xi \phi)^2. \qquad (20)$$

The first term on the r.h.s. of formula (2) gives just the energy of the background on the segment occupied by the kink. The second term is equal to the proper energy of the kink. It can be regarded as its rest mass $M$. We have computed $E$ and $M$ numerically. Sample results are given in the Table below.

| $\phi_0$ | $E$ | $M$ | $\xi_0$ |
|---|---|---|---|
| 1.5 | -0.263 | 3.128 | 1.824 |
| 1.0 | -0.062 | 1.478 | 1.674 |
| 0.2 | -0.019 | 0.044 | 1.574 |
| 0.05 | -0.002 | 0.002 | 1.570 |

Table 1. Results of numerical computations of the total energy of the field and of the rest mass of the compacton.

We see that the total energy $E$ of the field is negative in spite of the presence of the kink with the positive rest mass. We have checked that this is the case also for other choices of $\phi_0$.

## 3 The mechanical system with the topological compacton

Let us take a thick rectilinear wire with $2N+1$ pendulums connected to it at the points $x_i$, $i = -N, ..., N$. The points $x_i$ are separated by a constant distance $a$. Each pendulum has a very light arm of length $R$ and a mass $m$ at the free end. It can swing only in the plane perpendicular to the wire. All pendulums are fastened to the wire stiffly, hence their swings twist the wire accordingly. The wire is elastic with respect to such twists. Each pendulum has one degree of freedom which may be represented by the angle $\Phi(x_i, t)$ between the vertical direction and the pendulum. Thus, $\Phi(x_i, t) = 0$ corresponds to the upward position of the $i$-th pendulum. When all pendulums point upwards the wire is not twisted.

Equation of motion for each pendulum, except for the first and the last ones, has the following form

$$mR^2 \frac{d^2 \Phi(x_i, t)}{dt^2} = mgR \sin \Phi(x_i, t) + \kappa \frac{\Phi(x_i - a, t) + \Phi(x_i + a, t) - 2\Phi(x_i, t)}{a}, \quad (21)$$

where $\kappa$ is a constant which characterizes the elasticity of the wire with respect to the twisting. The first term on the r.h.s. of Eq.(21) is due to the gravitational force acting on the mass $m$, the second term is the elastic torque due to the torsion of the wire.

As for the two outermost pendulums, we assume that they are kept in the upward position by some external force, that is that

$$\Phi(x_{-N}, t) = 0, \quad \Phi(x_N, t) = 0. \tag{22}$$

The mechanical system described above is esssentially identical with one used in a realization of sinus-Gordon solitons, except that in the case of these solitons all pendulums initially point downwards, so that they are in the stable equilibrium position. In the present case they are put into the seemingly unstable equilibrium position $\Phi(x_i, t = 0) = 0$. However, as we show below, this position can actually be stable due to the presence of the external force which implements conditions (22). Another difference with the sinus-Gordon case is that we mechanically restrict the range of $\Phi(x_i, t)$

$$|\Phi(x_i, t)| \leq \Phi_0 < \pi, \tag{23}$$

by putting on both sides of the wire and parallely to it two rigid rods. The pendulums rebound from the rods if $\Phi(x_i, t) = \pm \Phi_0$.

Let us introduce a function $\Phi(x, t)$, where $x$ is a real continuous variable, twice differentiable with respect to $x$ and such that its values at the points $x = x_i$ are equal to the angles $\Phi(x_i, t)$. Hence, $\Phi(x, t)$ interpolates between $\Phi(x_i, t)$. Of course, for a given set of values of the angles there is an infinite number of such functions. For any of them the following identity holds

$$\Phi(x_i - a, t) + \Phi(x_i + a, t) - 2\Phi(x_i, t) = \int_0^a ds_1 \int_{-a}^0 ds_2 \frac{\partial^2 \Phi(s_1 + s_2 + x, t)}{\partial x^2}\bigg|_{x=x_i}.$$

We shall restrict our considerations to such motions of the pendulums that there exists the interpolating function $\Phi(x, t)$ of continuous variables $x, t$ such that

$$\int_0^a ds_1 \int_{-a}^0 ds_2 \frac{\partial^2 \Phi(s_1 + s_2 + x, t)}{\partial x^2}\bigg|_{x=x_i} \approx a^2 \frac{\partial^2 \Phi(x, t)}{\partial x^2}\bigg|_{x=x_i} \tag{24}$$

for all times $t$ and at all points $x_i$. For example, this is the case when the second derivative of $\Phi$ with respect to $x$ is almost constant when $x$ runs through the interval $[x_i - a, x_i + a]$, for all times $t$. With the approximation (24) the identity written above can be replaced by the following approximate one

$$\Phi(x_i - a, t) + \Phi(x_i + a, t) - 2\Phi(x_i, t) \approx a^2 \frac{\partial^2 \Phi(x, t)}{\partial x^2}\bigg|_{x=x_i}.$$

Using this formula in Eq.(21) we obtain

$$mR^2 \frac{d^2 \Phi(x_i, t)}{dt^2} \approx mgR \sin \Phi(x_i, t) + \kappa a \frac{\partial^2 \Phi(x, t)}{\partial x^2}\bigg|_{x=x_i}. \tag{25}$$

Let us now suppose that our function $\Phi(x,t)$ obeys the following partial differential equation,

$$mR^2\frac{\partial^2\Phi(x,t)}{\partial t^2} = mgR\sin\Phi(x,t) + \kappa a\frac{\partial^2\Phi(x,t)}{\partial x^2}, \qquad (26)$$

where $x \in [-Na, Na]$, and

$$\phi(-Na, t) = 0, \quad \phi(Na, t) = 0. \qquad (27)$$

Then, it is clear that $\Phi(x_i, t)$ obey Eq.(25) and the boundary conditions (22). Hence, if condition (24) is satisfied we obtain the approximate solution of the initial Newton equations (21). In the final step, we pass to the dimensionless variables

$$\tau = \sqrt{\frac{g}{R}}t, \quad \xi = \sqrt{\frac{mgR}{\kappa a}}x, \quad \phi(\xi, \tau) = \Phi(x, t).$$

Then, Eq.(26) acquires the form

$$\frac{\partial^2\phi(\xi,\tau)}{\partial\tau^2} - \frac{\partial^2\phi(\xi,\tau)}{\partial\xi^2} - \sin\phi(\xi,\tau) = 0. \qquad (28)$$

This equation coincides with the Euler-Lagrange equation obtained from Lagrangian (1) with potential (12) for $|\phi| < \phi_0$. The restriction (23) now has the form

$$|\phi(\xi,\tau)| \leq \phi_0.$$

The range of the spatial coordinate $\xi$ is from $-\xi_N$ to $\xi_N$, where $\xi_N = N\sqrt{mgRa/\kappa}$. The boundary conditions for $\phi$ have the form

$$\phi(-\xi_N, \tau) = 0 = \phi(\xi_N, \tau). \qquad (29)$$

Equation (28) with boundary conditions (29) has the trivial solution

$$\phi(\xi, \tau) = 0. \qquad (30)$$

Let us check stability of this solution against small perturbations. To this end we write $\phi(\xi, \tau) = \epsilon(\xi, \tau)$ and we linearize Eq.(28) with respect to $\epsilon$. The resulting equation has the form

$$\frac{\partial^2\epsilon}{\partial\tau^2} - \frac{\partial^2\epsilon}{\partial\xi^2} - \epsilon = 0.$$

It leads to the following dispersion relation

$$\omega(k) = \pm\sqrt{k^2 - 1}, \quad k = \frac{\pi}{2\xi_N}n,$$

where $n$ is an integer different from zero. The discrete values of $k$ are due to the boundary conditions (29). We see that the trivial solution is stable if $\xi_N < \pi/2$. Loosely speaking, this condition is satisfied when the wire is short enough.

The kink can appear only if the trivial solution is not stable. Then the boundary conditions (29) can be satisfied by locating one half of the kink at the one end of the wire and one half of the anti-kink at the other end. Each of them has the length $\xi_0$, where $\xi_0$ has been introduced in previous Section. We have to put at least one full kink in the middle. Therefore, the total dimensionless length of the wire has to be equal at least to $4\xi_0$. Thus, full compactons can be observed if

$$\xi_N > 2\xi_0.$$

Finally, let us note that we made a mechanical model of that kind, and the compactons were clearly seen.

## 4 Remarks

1. The single kink solution discussed in Section 2 can be trivially generalized. Lagrangian (1) and condition (13) are Poincaré invariant. Therefore, Lorentzian boosts give kinks moving with arbitrary velocities not exceeding 1 in the dimensionless variables. Because the kinks do not feel each other when their centers are separated by a distance larger that $2\xi_0$, it is clear that there exist static multi-kink solutions. Such 'trains' of separated kinks and anti-kinks are also the compacton type solutions. Moreover, also kinks and anti-kinks moving with different velocities can be trivially combined to give compacton solutions in a finite time interval until they touch each other.

2. Our work can be extended in several directions. One could study the related non-planar domain walls. We have mentioned in the Introduction that they can be constructed with the help of expansion in width.

Another direction is suggested by the fact that the separated compactons do not interact. Therefore, they seem to be well-suited for testing theoretical descriptions of productions of topological defects, [2, 3, 4]. The lack of interactions in the final state simplifies counting of the defects. Our current work is devoted to this topic.

One could also study interaction of the kink with the anti-kink, for example, when they scatter on each other with various relative velocities.

# 5 Acknowledgement

We would like to thank Dr. P. Węgrzyn for remarks pertinent to this work.